\documentstyle[titlepage,12pt]{article}
\def\bc{\begin{center}}
\def\ec{\end{center}}

\def\beq{\begin{equation}}
\def\eeq{\end{equation}}
\def\barr{\begin{eqnarray}}
\def\earr{\end{eqnarray}}
\def\sech{{\rm sech}}
\def\tanh{{\rm tanh}}
\newfont{\fonta}{msbm10 scaled\magstep1}
\begin{document}
\begin{titlepage}
\renewcommand{\thefootnote}{\fnsymbol{footnote}}
\begin{flushright}
TMUP-HEL-9406\\
May, 1994
\end{flushright}
\vskip 1.0cm
\begin{center}
{\bf Solitons in Nonlinear Schr\"{o}dinger Model and
the Collective Ground State of One-Dimensional Delta-Function Gas}
\vglue 0.5cm
\end{center}
\vskip 0.5cm
\begin{center}
Hideaki Hiro-Oka\footnote[1]{Research Fellow of Japan Society for the
			     Promotion of Science}%
                \footnote[2]{e-mail: hiro-oka@phys.metro-u.ac.jp} and %
        Hisakazu Minakata\footnote[3]{e-mail: minakata@phys.metro-u.ac.jp}
\end{center}
\vskip 1.0cm
\begin{center}
{\it Department of Physics\\
Tokyo Metropolitan University\\
1-1 Minami-Osawa Hachioji, Tokyo 192-03, Japan}
\end{center}
\vskip 3cm
\begin{flushleft}
{\bf Abstract}
\end{flushleft}

We examine one-dimensional Bose gas interacting with delta-function
potential using the large-$N$ collective field theory. We show that in
the case of attractive potential the uniform ground state is unstable to
small perturbations and the instability is cured by formation of a collective
ground state, \lq\lq bright soliton'' configuration in corresponding
nonlinear Schr\"odinger field theory.

\end{titlepage}
\noindent

The large-$N$ collective field theory formulated by Jevicki and Sakita
\cite{Jevicki}\cite{Sakita} has been known as a powerful nonperturbative tool
in dealing with
many-body collective phenomena.
It is remarkable that its applicability extends over wide range of problems
such as, plasma oscillations \cite{Jevicki}\cite{Sakita}, matrix models
\cite{Jevicki}\nocite{Sakita}-\cite{Andric}, the quantum
Hall effect \cite{Sakita2}, and anyons \cite{Hiro-Oka}.

In this paper we demonstrate that the large-$N$ collective field theory
is a really powerful machinery in uncovering collective modes at
high densities. We shall do this in a particular model of bosons
in 1 dimension interacting via a delta-function potential.
Despite that our discussion will be restricted to this particular model
we believe that the power of
the formalism in hunting collective excitations extends over a wide class
of models.

One-dimensional Bose gas with delta-function potential have been
known to be soluble \cite{Lieb} since long time ago.
Nonetheless, it provides a nontrivial testing ground in the context
of the collective field formalism and we will learn some
meaningful lessons.
First of all we will verify that the formalism reproduces the results
obtained by Lieb and Liniger \cite{Lieb} in high-density limit. We
observe, not surprisingly, that the ground state energy in the case of
repulsive potential
is of the order of $\hbar^0$ and is in agreement with that of
ref.\cite{Lieb}.
We also obtain a classical sound velocity consistent with the classical
ground state energy. More importantly, we clarify how the instability
of the uniform ground state in the case of attractive potential can be
stabilized by forming a collective ground state, a \lq\lq bright soliton''
configuration of the corresponding nonlinear Schr\"odinger field theory.
We also give some remarks on the role of \lq\lq dark soliton''
configurations in the case of repulsive potential.

The system of $N$ particles interacting with a delta-function potential
is described by the Hamiltonian
\begin{equation}
H = \displaystyle -\frac{\hbar^2}{2m} \sum^N_{a=1}
\frac{{\partial}^2}{\partial x_a^2} + 2c \sum_{a>b} \delta\: (x_a-x_b).
\label{eqn:Hamiltonian}
\end{equation}
Our notation is the same as that of Lieb and Liniger \cite{Lieb} except
that they take $m=1/2$.
Also we shall explicitly denote $\hbar$ throughout this paper for
convenience of our discussions.
The canonical approach to the problem is, of course, to solve the
eigenvalue problem with the Hamiltonian (\ref{eqn:Hamiltonian}) using
a symmetric (Bose) and an antisymmetric (Fermi) wave functions.
This is beautifully done by Lieb and Liniger in the former case
by using the Bethe ansatz technique.

Instead, we shall develop a collective field theory treatment of this
system in this paper. Following the well-established way [2]
we rewrite the Hamiltonian (\ref{eqn:Hamiltonian}) in terms of
the density variables
\begin{equation}
\phi(x) = \displaystyle \frac{1}{N} \sum^N_{a=1} \delta\:(x-x_a)
\label{eqn:density}
\end{equation}
and its canonical conjugate $\pi(x)$. Here the density variable
is normalized such that
\beq
\int dx\phi(x) =1.\label{normalization}
\eeq
The comments that follow are well-known but are in order
for readers who are unfamiliar with
the large-$N$ collective field theory: One should notice
that the Hamiltonian we shall work with is not the one
which we will obtain by just changing the variables from $x_a$ to $\phi(x)$.
Let us denote the Jacobian as $J$ of the transformation
of $x_a$ to $\phi(x)$ in (\ref{eqn:density}),
or more precisely speaking, to its Fourier transformation
\begin{eqnarray}
\phi_k &=& \displaystyle \int dx \mbox{e}^{-ikx} \phi(x)\nonumber\\
       &=& \displaystyle \frac{1}{N} \sum^N_{a=1} \mbox{e}^{-ikx_a}.
\label{eqn:transform}
\end{eqnarray}
Then the functional space we are to work with is that of
Schr\"{o}dinger's wave function $J^{1/2}\Psi$ not $\Psi$,
where $\Psi$ denotes the Schr\"{o}dinger's wave function of the original
Hamiltonian (\ref{eqn:Hamiltonian}).
Therefore, the Hamiltonian for our system written in terms of
$\phi(x)$ is $J^{1/2}H(\phi,\pi)J^{-1/2}$, and this is the Hamiltonian
which possesses hermiticity in the functional space of $J^{1/2}\Psi$.
We shall denote this new Hamiltonian as $H_{eff}$.

We obtain $H_{eff}$ in the following form:
\begin{eqnarray}
H_{eff} &=& \displaystyle \frac{\hbar^2}{2mN}\int dx
(\partial\pi)\phi(\partial\pi)
         + \frac{\hbar^2N}{8m}\int dx \frac{(\partial\phi)^2}{\phi}
\nonumber\\
        & & + N^2c \int dx \phi^2
         - \lambda N \left(\int dx\phi - 1\right),
\label{eqn:Heff}
\end{eqnarray}
where $\partial$ is an abbreviation of
$\displaystyle \frac{\partial}{\partial x}$.
This is the Hamiltonian which we analyze using the large-$N$ technology.
Note that we only have a nontrivial interacting theory
under the large-$N$ limit in which $cN$ is finite and $N\rightarrow\infty$.
In this case all the terms in (\ref{eqn:Heff}) are of order $N$
assuming that $\pi$ is of order $N$.

Let us discuss the ground state and the low-lying excited states of
the system (\ref{eqn:Heff}). Since the Hamiltonian is of order $N$
we may allow to use saddle-point approximation, e.g.,
in the path integral formalism. As a result we have the classical
Hamilton's equations.
For time-independent ground-state configuration they read,
%
\beq
\partial(\phi \partial \pi) = 0,\label{eqn:ground}
\eeq
\begin{equation}
\label{eqn:state}
\displaystyle
\frac{\hbar^2}{2m}(\partial\pi)^2-\frac{\hbar^2}{4m}\frac{{\partial}^2\phi}{\phi}
 + \frac{\hbar^2}{8m}\left(\frac{\partial\phi}{\phi}\right)^2 + 2Nc\phi - %
\lambda = 0.
\end{equation}
We require that the ground state is translationally invariant.
Then the solution is, rather trivially,
\begin{equation}
\phi_0(x) = \displaystyle \frac{1}{L}, \quad \pi_0(x) = 0, \label{eqn:solution}
\end{equation}
where we have put the system into the box of length $L$.
By virtue of (\ref{eqn:state}) we have the relation between
the chemical potential $\lambda$ and
the density of particles $\rho=N/L$ as $\lambda=2c\rho$.
Using (\ref{eqn:Heff}) and (\ref{eqn:solution})
we obtain the ground state energy of the system as
\begin{eqnarray}
E_0 &=& N^2 c \cdot \displaystyle \frac{1}{L},\nonumber\\
    &=& Nc\rho,
\label{eqn:system}
\end{eqnarray}
in agreement with the result of Lieb-Liniger \cite{Lieb} at high-density
($\rho\rightarrow\infty$) limit in the repulsive potential $c>0$.
Thus our machinery works in the large-$N$ limit.
We note that the ground state energy (\ref{eqn:system}) is a classical one,
namely, it is free from $\hbar$. This feature
is shared by the exact result obtained in ref.\cite{Lieb}
as one can verify by simple dimensional analysis.
While somewhat curious at first site it is an example of
the well-known fact that the system becomes classical in high
density limit.

So far our discussion is valid not only for repulsive $(c>0)$ potential
but also for attractive $(c<0)$ potential. Apparently the ground state
(\ref{eqn:solution}) exists also for the latter case in contradiction to the
claim given in ref.\cite{Lieb}.

To resolve the problem and to verify the validity of the large-$N$ collective
field theory we examine low-lying excitations as small fluctuations
around the ground state (\ref{eqn:solution}).
We employ plane-wave basis to expand the fluctuations as
\begin{eqnarray}
\phi(x) &=& {1\over L}\left(1+\sum_k \sqrt{k^2\over N}e^{ikx}Q_k\right),
\nonumber\\
\pi(x) &=& {1\over \hbar}\sum_k \sqrt{N\over k^2}e^{-ikx}P_k.
\label{eqn:fluctuations}
\end{eqnarray}
Inserting (\ref{eqn:fluctuations}) into the Hamiltonian
(\ref{eqn:Heff}) and picking up the order $N^0$ term
[ order $N^{1/2}$ term cancels because of equations of motion
(\ref{eqn:ground}) and (\ref{eqn:state})]
we obtain the Hamiltonian of harmonic oscillator form
\begin{equation}
\label{eqn:harmonic}
H_{N^0}=\sum_k\left(\displaystyle \frac{1}{2m}|P_k|^2
+ \frac{m}{2}\omega^2(k)|Q_k|^2\right),
\end{equation}
where
\begin{equation}
\label{eqn:oscillator}
\hbar^2\omega^2(k) = \left(\displaystyle \frac{\hbar^2}{2m}k^2\right)^2
+ \frac{2c\rho}{m}\hbar^2k^2.
\end{equation}
In the case of repulsive potential each term of (\ref{eqn:oscillator})
allows a simple physical interpretation:
the first term is an obvious kinetic energy term, and the second one is
the energy of sound wave with the velocity
\begin{equation}
\label{eqn:sound}
v_s=\displaystyle \sqrt\frac{2c\rho}{m}.
\end{equation}
By a well-known macroscopic argument the sound velocity is given by
\begin{equation}
\label{eqn:velocity}
v_s = \sqrt{-\frac{L}{m\rho}\frac{\partial P}{\partial L}}
\end{equation}
where $P$ is the pressure,
\begin{equation}
\label{eqn:pressure}
P = \displaystyle -\frac{\partial E_0}{\partial L}.
\end{equation}
It is amusing to observe that the use of (\ref{eqn:velocity}),
(\ref{eqn:pressure}),
and (\ref{eqn:system}) reproduces the expression (\ref{eqn:sound}) for the
sound
velocity. This is, of course, nothing but
repeating the argument in ref.\cite{Lieb} under
the high-density approximation.
We just want to point out that the sound velocity (\ref{eqn:sound}) has
a classical value because the ground state energy (\ref{eqn:system}) is
classical.
Thus we have seen that the large-$N$ collective field theory works for Bose gas
with repulsive delta-function potential in the high-density limit.

In the case
of attractive potential, $c<0$, the interpretation of (\ref{eqn:oscillator})
is quite different. At sufficiently small $k$ the frequency $\omega$ becomes
imaginary signaling the instability of the ground state (\ref{eqn:solution}).
In this way we have reached a conclusion that there is no stable {\it uniform}
ground state for attractive potential, in agreement with ref.\cite{Lieb}.

Now we turn to the discussion of collective ground state. Namely, we show that
there exist solutions in (\ref{eqn:ground}) and (\ref{eqn:state}) which can
be interpreted as collective excitation
of Bose particles at high-densities.

To this goal we introduce a Schr\"{o}dinger variable
\beq
\psi(x) = \sqrt{\phi(x)} \mbox{e}^{-i\pi(x)}.\label{Schroedinger}
\eeq
Notice that there is no direct connection between this $\psi$ and
the many-body Schr\"{o}dinger wave function of the original problem.
Then it is easy to show that the equations (\ref{eqn:ground})
and (\ref{eqn:state})
can be rewritten as
\begin{equation}
-\displaystyle \frac{\hbar^2}{2m}\frac{d^2}{dx^2}\psi(x)
 + 2Nc|\psi|^2\psi(x) = \lambda\psi(x)
\label{eqn:rewrite}
\end{equation}
This is nothing but the classical nonlinear Schr\"{o}dinger
equation.

It is well known that the equation (\ref{eqn:rewrite}) admits
an infinite number of soliton solutions \cite{ZS}.
Solitons in theories with attractive and repulsive interactions are
called \lq\lq bright'' and \lq\lq dark'' solitons, respectively.
These names originates in its application to the problem of
nonlinear optical pulses in dispersive dielectric fibers \cite{Hasegawa}.

As discussed before there exists no stable uniform ground state for $c<0$
case. However, we see that the collective ground state can exist, which
corresponds to the bright soliton. In the case of repulsive interaction
we can construct the dark soliton solution besides the uniform ground
state.

For the purpose of illuminating the general picture we work with a simple
ansatz of the bright soliton solution of the form
\beq
\psi(x)\propto \sech\ ax.
\eeq
In our language we assume
\beq
\phi=A\ \sech^2 ax,\quad\pi=B,\label{bright}
\eeq
from the relation (\ref{Schroedinger}). The unknown constants $a$, $A$,
and $B$ are determined by the requirement that (\ref{bright}) are solutions
to (\ref{eqn:ground}) and (\ref{eqn:state}). Substituting (\ref{bright}) into
(\ref{eqn:ground}) and (\ref{eqn:state}) we obtain
\beq
\left({\hbar^2a^2\over m}+2NcA\right)\ \sech^2 ax-
\left(\lambda+{\hbar^2a^2\over{2m}}\right)
=0.\label{relation}
\eeq
This relation lead to
\beq
A=-{{\hbar^2a^2}\over{2Ncm}},\quad \lambda=-{{\hbar^2a^2}\over{2m}},%
\label{result1}
\eeq
apart from a trivial case $a=\lambda=0$.
By definition $\phi$ and $A$ must be positive,
so we have two cases: i) $c<0$ and $a\in$ {\fonta R}, ii) $c>0$ and
$a=i\alpha$,
$\alpha\in$ {\fonta R}. As can be seen the case ii) yields the unphysical
solution with singularities. Thus we have
\beq
\phi={a^2\over{2N{\tilde c}m}}\ \sech^2 ax,\label{nresult1}
\eeq
where ${\tilde c}=-c>0$. On the other hand $\pi$ is an arbitrary constant.
Due to the normalization condition for $\phi$, $a$ is determined
as
\beq
a={N{\tilde c}m\over \hbar^2}.\label{coupling}
\eeq
This is the collective ground state. We obtain the ground state energy as
\beq
E_0=-{m\over {6\hbar^2}}N(Nc)^2,\label{bgenergy}
\eeq
by inserting (\ref{nresult1}) to the effective Hamiltonian.
Note that this energy is in agreement
with that of a static bright soliton solution of the classical nonlinear
Schr\"odinger problem.

Next we consider the dark soliton solution of the form
\beq
\phi=A\ \tanh^2 ax,\quad\pi=B,\label{dark}
\eeq
in the similar way. Note that this is a special case of generic dark
soliton solutions.
Substituting (\ref{dark}) into (\ref{eqn:ground})
and (\ref{eqn:state}) we obtain
\beq
\left(2NcA-{{\hbar^2a^2}\over m}\right)\ \tanh^2 ax+%
\left({{\hbar^2a^2}\over m}-\lambda\right)
=0.\label{relationdark}
\eeq
By the same discussion we have
\beq
A={{\hbar^2a^2}\over{2Ncm}},\quad\lambda={{\hbar^2a^2}\over m}.%
\label{resultdark}
\eeq
The constant $a$ is given by the normalization of $\phi$,
\beq
|a|={1\over L}\left(1+\sqrt{1+{{2Ncm}\over \hbar^2}L}\right).
\eeq
The ground state energy is given as
\beq
E_0=N\rho c+{4\over 3}{{\hbar\rho}\over m}\sqrt{2m\rho c}%
\label{dgenergy}
\eeq
apart from the correction of the order of $e^{-L}$.
The first term of (\ref{dgenergy}) is nothing but the energy of
the uniform ground state while the second term is the quantum correction of
order $\hbar$.

To summarize we make a few remarks:

i) We have clarified, within the framework
of large-$N$ collective field theory, how the instability of the uniform
ground state of the attractive delta-function gas can be cured by the
formation of the collective ground state. The feature that the ground state
itself is a non-uniform collective state seems to be very new to us.

ii) In the case of repulsive potential the ground state is uniform and is
stable against small perturbations. We have reproduced Lieb-Liniger's result
on the energy of the fluctuation mode in high-density limit. We also have
dark soliton excitation on this ground state.

iii) The quantum meaning of the bright soliton solution of the classical
nonlinear Schr\"odinger field theory has been thoroughly discussed by Wadati
and coworkers \cite{Wadati}. The large-$N$ collective field formalism can
be viewed as a way of constructing quantum field theory starting from $N$-%
body Schr\"odinger problem. This is achieved by taking the thermodynamic
limit, $N\rightarrow \infty$ keeping $N/L$ fixed. Whereas we have discussed
the high-density limit, $N\rightarrow\infty$ keeping $L$ large but fixed in
this paper and the clear relationship with classical soliton solutions has
emerged. Therefore, the large-$N$ collective theory provides an another way
of understanding quantum correspondence with the classical object.

Works of H. M. and H. H. are partially supported by Grant-in-Aid for
Scientific Research of the Ministry of Education, Science and Culture
\#05640355 and \#H4-2028, respectively.

\vglue 1cm

\end{document}